\documentclass[11pt]{article}

\usepackage[
citestyle=authoryear,
style=authoryear,
uniquename=false,
sorting=ynt,
maxbibnames=99
]{biblatex}

\usepackage[font=small,labelfont=bf]{caption}

\usepackage{enumitem}
\usepackage{graphicx}
\usepackage{xcolor}
\usepackage{blkarray}
\usepackage{amsmath,amssymb}

\begin{document}
\setlength{\abovedisplayskip}{4pt}
\setlength{\belowdisplayskip}{10pt}
\setlength{\abovedisplayshortskip}{4pt}
\setlength{\belowdisplayshortskip}{10pt}

\title{An activation-clearance model for \textit{Plasmodium vivax} malaria}
\author{Somya Mehra$^1$ \and James M. McCaw$^{1,2,3}$ \and Mark B. Flegg$^4$ \and Peter G. Taylor$^1$ \and Jennifer A. Flegg$^1$}

\date{%
    $^1$School of Mathematics and Statistics, The University of Melbourne, Melbourne, Australia\\%
    $^2$Centre for Epidemiology and Biostatistics, Melbourne School of Population and Global Health, The University of Melbourne, Melbourne, Australia\\
    $^3$Peter Doherty Institute for Infection and Immunity, The Royal Melbourne Hospital and The University of Melbourne, Australia, Parkville, Victoria, Australia \\
    $^4$School of Mathematics, Monash University, Melbourne, Australia\\[2ex]%
}
\maketitle


\section{Abstract}
Malaria is an infectious disease with an immense global health burden. \textit{Plasmodium vivax} is the most geographically widespread species of malaria. Relapsing infections, caused by the activation of liver-stage parasites known as hypnozoites, are a critical feature of the epidemiology of \textit{Plasmodium vivax}. Hypnozoites remain dormant in the liver for weeks or months after inoculation, but cause relapsing infections upon activation. Here, we introduce a dynamic probability model of the activation-clearance process governing both potential relapses and the size of the hypnozoite reservoir. We begin by modelling activation-clearance dynamics for a single hypnozoite using a continuous-time Markov chain. We then extend our analysis to consider activation-clearance dynamics for a single mosquito bite, which can simultaneously establish multiple hypnozoites, under the assumption of independent hypnozoite behaviour. We derive analytic expressions for the time to first relapse and the time to hypnozoite clearance for mosquito bites establishing variable numbers of hypnozoites, both of which are quantities of epidemiological significance. Our results extend those in the literature, which were limited due to an assumption of non-independence. Our within-host model can be embedded readily in multi-scale models and epidemiological frameworks, with analytic solutions increasing the tractability of statistical inference and analysis. Our work therefore provides a foundation for further work on immune development and epidemiological-scale analysis, both of which are important for achieving the goal of malaria elimination.\\

\section{Introduction}
Approximately 2.5 billion people globally are at risk of developing malaria \parencite{howes2016global}. The global malaria burden is immense, with an estimated 216 million cases and 445,000 deaths globally in 2016 \parencite{who2017}. The primary pathogens responsible for human malaria are \textit{Plasmodium falciparum} and \textit{Plasmodium vivax}. \textit{Plasmodium vivax} contributes to a significant proportion of the malaria burden beyond sub-Saharan Africa, causing 64\% of malaria cases in the Americas, 30\% in South-East Asia and 40\% in the Eastern Mediterranean \parencite{who2017}.\\

Relapsing infections are a critical feature of the epidemiology of \textit{Plasmodium vivax}. The bite of an infected \textit{Anopheles} mosquito may lead to a primary blood infection, as well as the establishment of dormant parasite stages known as hypnozoites. Hypnozoites remain dormant within liver cells (hepatocytes) for indeterminate periods, but may cause further blood-stage infection (relapse) upon activation \parencite{mueller2009key}. The clearance of hypnozoites from a human host can be achieved either through activation, or the death of hypnozoite,  possibly due to the death of the host hepatocyte \parencite{malatohep}. Both infected \textit{Anopheles} mosquito vectors and reservoirs of dormant hypnozoites contribute to the force of infection for \textit{Plasmodium vivax}, further adding to the difficulty of disease control and elimination \parencite{howes2016global, mueller2009key}.\\

The number of relapses arising from a mosquito bite is variable and thought to be dependent on the size of the initial parasite inoculum, host immunity and the inoculated strain of \textit{Plasmodium vivax} parasites \parencite{white2011determinants}. The mechanisms of hypnozoite activation are poorly understood, with hypothesised triggers including systemic febrile illness \parencite{whiteimwong} and bites from mosquitoes \parencite{hulden}. Here we aim to capture a ‘baseline scenario’, modelling hypnozoite activation to occur at a constant rate post-dormancy without accounting for possible external triggers such as febrile illness and mosquito bites \parencite{whitemodel}. We restrict our analysis to hypnozoite activation events, without considering the effects of immunity on observed relapses.\\

An existing long-latency model of the hypnozoite reservoir, including an explicit dormancy stage during which activation cannot occur, has been developed by \textcite{whitemodel}. White's model, however, enforces a ``collective dormancy" for hypnozoites established through the same mosquito bite; hypnozoites from the same mosquito bite progress through compartments in the latency phase in lock-step, and become susceptible to activation at the same instant. \\

Collective dormancy could be feasible if hypnozoites from the same mosquito bite inhabited the same host liver cell (hepatocyte). Given the complex trajectory of infected sporozoites through the skin and bloodstream before reaching the host liver, and the subsequent transmigration through multiple hepatocytes \parencite{rankinimaging}, it is highly unlikely for hypnozoites established through the same mosquito bite to inhabit the same host liver cell. The assumption of collective dormancy, moreover, would necessitate coupling in the death rates of hypnozoites, since hypnozoite death can be caused by the death of the host liver cell \parencite{malatohep}; however, in White's model, each hypnozoite is assumed to die independently at a constant rate during the latency phase \parencite{whitemodel}.\\

In this paper, we develop a within-host model of hypnozoite activation to examine the dynamics of relapsing infections, building on the model of the hypnozoite reservoir developed by \textcite{whitemodel}. We adapt their long-latency model to allow hypnozoites to proceed through the latency phase independently. Our model is not predicated on an underlying interaction between hypnozoites established through the same mosquito bite, and treats each hypnozoite to be statistically independent. Our framework allows us to examine relapses arising from the activation of hypnozoites, in addition to the clearance of the hypnozoite reservoir.  \\

The paper is structured as follows. Section \ref{single} examines the activation-clearance of a single hypnozoite. In Section \ref{multiple}, we extend our analysis to consider hypnozoites established through the same mosquito bite, deriving metrics including time to first relapse and the duration of hypnozoite carriage. In Section \ref{comparison} we give a comparative analysis illustrating how the assumption of independent hypnozoite behaviour gives rise to different dynamics for our model relative to White's model before making concluding remarks in Section \ref{discussion}.\\

\section{Modelling the dynamics of a single hypnozoite}\label{single}
\subsection{Model development}
\begin{figure}[ht]
\centering
\includegraphics[width=0.75\textwidth]{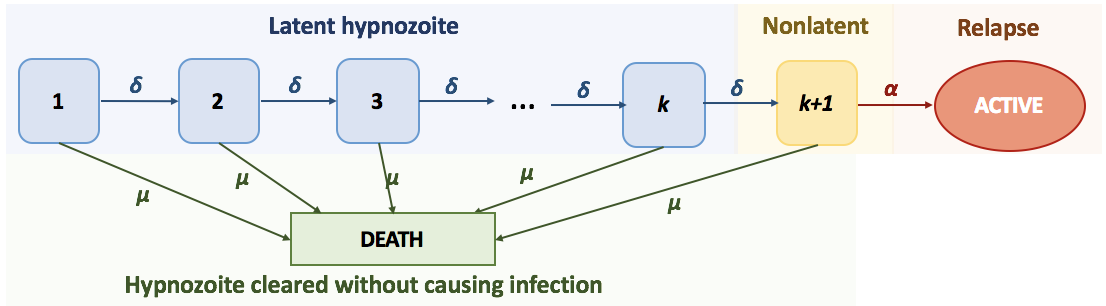}
\caption{Schematic for model of dynamics for a single hypnozoite. Hypnozoites progress through $k$ compartments of the latency phase becoming becoming nonlatent. $\delta$ is the rate at which hypnozoites progress through latency compartments, $\mu$ is the rate at which both latent and nonlatent hypnozoites die, and $\alpha$ is the rate at which nonlatent hypnozoites activate.}
\label{hypnozoite_activation}
\end{figure}

We first consider the dynamics of a single hypnozoite. There are two absorbing states: death and activation. Before reaching an absorbing state, a hypnozoite may either be latent, in which case it can die but not activate, or nonlatent, in which case it can either die or activate. We assume that each hypnozoite undergoes a latency phase before it is able to activate. We model the state of a hypnozoite using a continuous-time Markov chain with $(k+1)$ non-absorbing compartments and absorbing compartments ``death" and ``active", as shown in Figure \ref{hypnozoite_activation}. The first $k$ compartments constitute the latency phase of the hypnozoite, during which it may either progress to the subsequent compartment at constant rate $\delta$ or die at a rate constant rate $\mu$, possibly due to the death of the host liver cell. In compartment $(k+1)$, we consider a hypnozoite to be nonlatent. Here, it may either activate at constant rate $\alpha$ or die at constant rate $\mu$.

Consider a single hypnozoite inoculated at time zero. Let $X(t)$ denote the state of the hypnozoite at time $t$. The probability mass function (PMF) of $X(t)$ is given by
\begin{align*}
    \mathbf{p}(t) = \big( p_{1}(t), p_{2}(t), \dots, p_{k}(t), p_{\text{nonlatent}}(t), p_{\text{active}}(t), p_{\text{death}}(t) \big) ,
\end{align*}
where we define the state probabilities
\begin{itemize}
    \item $p_{m}(t)$ that a hypnozoite is in compartment $m \in [1, k]$ of the latency phase,
    \item $p_{\text{nonlatent}}(t)$ that a hypnozoite is nonlatent, that is, in compartment $k+1$,
    \item $p_{\text{active}}(t)$ that a hypnozoite has activated, and
    \item $p_{\text{death}}(t)$ that hypnozoite death has occurred.\\
\end{itemize}

Using the transition rates shown in Figure \ref{hypnozoite_activation}, we obtain the transition rate matrix
\begin{align}
\mathbf{Q} = 
\begin{blockarray}{c c c c c c c c c}
\textbf{1} & \textbf{2} & \textbf{3} & \textbf{\dots} & \textbf{k} & \textbf{\text{nonlat}} & \textbf{\text{active}} & \textbf{\text{death}} & \\
\begin{block}{[c c c c c c c c]c}
-(\delta+\mu) & \delta & 0 & \dots & 0 & 0 & 0 & \mu & \textbf{1} \\
0 & -(\delta+\mu) & \delta & \dots & 0 & 0 & 0 & \mu & \textbf{2} \\
\vdots & \vdots & \vdots & \ddots & \vdots & \vdots & \vdots & \vdots & \textbf{\vdots} \\
0 & 0 & 0 & \dots & -(\delta+\mu) & \delta & 0 & \mu & \textbf{k} \\
0 & 0 & 0 & \dots & 0 & -(\alpha + \mu) & \alpha & \mu & \textbf{\text{nonlat}} \\
0 & 0 & 0 & \dots & 0 & 0 & 0 & 0 & \textbf{\text{active}}\\
0 & 0 & 0 & \dots & 0 & 0 & 0 & 0 & \textbf{\text{death}} \\
\end{block}
\end{blockarray} \label{Q}
\end{align}
where $Q_{i,j}$ is the transition rate from state $i$ to state $j$. We assume each hypnozoite starts in the first compartment of the latency phase, yielding initial condition
\begin{align}
\mathbf{p}(0) = \mathbf{e_1} = \big( p_{1}(0), p_{2}(0), \dots, p_{k}(0), p_{\text{nonlatent}}(0), p_{\text{active}}(0), p_{\text{death}}(0) \big) = (1, 0, \dots, 0, 0, 0) \label{IC}
\end{align}

{{Kolmogorov Forward Differential Equations for the time evolution of PMF $\mathbf{p}(t)$ are}}
\begin{align}
    \frac{d \mathbf{p}(t)}{dt} = \mathbf{p}(t) \mathbf{Q}. \label{kolmogrov}
\end{align}
The system of $(k+3)$ ODEs resulting from the expansion of Equation (\ref{kolmogrov}) is provided in the Appendix.\\

The solution to Equation (\ref{kolmogrov}) {{ with}} initial condition (\ref{IC}) takes the form of the matrix exponential
\begin{align}
    \mathbf{p}(t) = \mathbf{e_1} e^{\mathbf{Q}t}. \label{matexp}
\end{align}
However, we can derive expressions for the components of ${\bf p}(t)$, as well as gaining physical insight into the problem, by considering the sequences of events that need to occur in the time interval $[0,t]$ for the process to reside in each of its compartments at time $t$.\\

To be in compartment $m \in \{1, \dots, k\}$ of the latency phase at time $t$, the Markov chain must have
\begin{itemize}
\item undergone exactly $(m-1)$ compartmental transitions, the probability of which is given by the Poisson PMF
\begin{equation}
\label{eq:poisson}
g_{\delta, m-1}(t) = \frac{(\delta t)^{m-1}}{(m-1)!}e^{-\delta t}
\end{equation} with rate parameter $\delta$, and
\item not been absorbed by death, which has probability $e^{-\mu t}$.
\end{itemize}

Since hypnozoite death and progression through the latency phase are modelled to be independent processes,
\begin{align}
	p_{m}(t) & = \frac{(\delta t)^{m-1}}{(m-1)!}e^{-(\mu+\delta)t} \text{ for } m \in [1, k]. \label{pm}
\end{align}

\vspace{4mm}

To be in the nonlatent state at time $t$, the Markov chain must have 
\begin{itemize}
\item moved through all the compartments of the hypnozoite state in some time $u \in [0,t]$, which has an Erlang density 
\begin{equation}
\label{eq:erlang}
f_{\delta,k-1}(u) = \frac{\delta^{k}u^{k-1}e^{-\delta u}}{(k-1)!}
\end{equation}
of order $k$ and parameter $\delta$,
\item not moved to the active state in the remaining time $t-u$, which has probability $e^{-\alpha(t-u)}$, and
\item not been absorbed by death, which has probability $e^{-\mu t}$.
\end{itemize}

Thus
\begin{align}
p_{\text{nonlatent}}(t) & = e^{-\mu t} \int_0^t \frac{\delta^{k}u^{k-1}e^{-\delta u}}{(k-1)!} e^{-\alpha(t-u)} du \label{nonlatentint} \\
&= \frac{\delta^{k}}{(\delta - \alpha)^{k}} \Bigg[ e^{-(\mu + \alpha)t} - e^{-(\mu + \delta)t} \sum^{k-1}_{j=0} \frac{t^j}{ j!} (\delta - \alpha)^{j} \Bigg] \label{nonlatent}
\end{align}
where we have evaluated Equation (\ref{nonlatentint}) using standard integral number 2.321.2 in \textcite{jeffrey2007table}, derived using integration by parts.\\

To be in the active state at time $t$, the Markov chain must have 
\begin{itemize}
\item moved through all the compartments of the hypnozoite state in some time $u \in [0,t]$, which has density given by (\ref{eq:erlang}), 
\item moved to the active state at some time $v \in [u,t]$ which has probability density $\alpha e^{-\alpha(v-u)}$, and
\item not been absorbed by death before time $v$, which has probability $e^{-\mu v}$.
\end{itemize}
Thus
\begin{align}
    p_{\text{active}}(t) = \int_0^t \int_u^t\frac{\delta^{k}u^{k-1}e^{-\delta u}}{(k-1)!} \alpha e^{-\alpha(v-u)} e^{-\mu v} dv du. \label{activeint}
\end{align}

Changing the order of integration in Equation (\ref{activeint}), we find
\begin{align}
    p_\text{{active}}(t) = \int_0^t \int_0^v\frac{\delta^{k}u^{k-1}e^{-\delta u}}{(k-1)!} \alpha e^{-\alpha(v-u)} e^{-\mu v} du dv = \int_0^t  \alpha p_\text{nonlatent}(v) dv.  \label{activeint2}
\end{align}

Similarly, we evaluate Equation (\ref{activeint2}) to yield 
\begin{align}
\begin{split}
    p_{\text{active}}(t) = & \frac{\alpha \delta^{k}}{(\delta - \alpha)^{k}} \Bigg[\frac{e^{-(\mu + \delta)t}}{\mu + \delta} \Bigg\{ \sum^{k-1}_{j=0} \Big( \frac{\delta - \alpha}{\mu + \delta} \Big)^j \sum^{j}_{i=0} \frac{t^i}{i!} (\mu + \delta)^{i} \Bigg\} - \frac{e^{-(\mu + \alpha)t}}{\mu + \alpha} \Bigg] +  \frac{\alpha}{\alpha + \mu} \Big( \frac{\delta}{\delta + \mu} \Big)^{k}.
	\label{pactive}
\end{split}
\end{align}

By the conservation of probability, the probability that the Markov chain will be in state death at time $t$ is therefore given by
\begin{align}
    p_\text{death}(t) = 1 - p_\text{active}(t) - p_\text{nonlatent}(t) - \sum^k_{m=1} p_m(t). \label{pdeath}
\end{align}

\subsection{Key quantitative results}
Since we have a Markov chain with constant transition rates, the probability $p_A$ that a hypnozoite will activate before dying is
\begin{align}
    p_A := \lim_{t \to \infty} p_{\text{active}}(t)  = \underbrace{\frac{\alpha}{\alpha+\mu}}_{\Pr[\text{activation} | \text{nonlatent}]} \times \underbrace{\Big(\frac{\delta}{\delta+\mu}\Big)^{k}.}_{\Pr[\text{survive $k$ latency compartments}]}\label{p_A}
\end{align}\

Therefore, given a hypnozoite does activate, the cumulative distribution function (CDF) for the activation time $A(t)$ is given by
\begin{align}
    A(t) = \Pr[\text{active at time } t | \text{activation occurs}] = \frac{p_{\text{active}}(t)}{p_A}, \label{A}
\end{align}
where $p_{\text{active}}(t)$ is the probability of activation time $t$ after inoculation, given by Equation (\ref{pactive}).\\

By definition, the expected time to relapse, $T_r$, given an activation does occur, is then
\begin{align}
    T_r = \int^{\infty}_{0} \Big( 1 - A(t) \Big) dt = \underbrace{\frac{k}{\mu+\delta}}_{\text{latency phase}} + \underbrace{\frac{1}{\alpha+\mu}.}_{\text{nonlatency phase}}
\end{align}

Using a similar argument, given a hypnozoite dies before activating, the CDF for the time of death is given by
\begin{align}
    D(t) = \frac{p_{\text{death}}(t)}{1-p_A},
\end{align}
where $p_{\text{death}}(t)$ is the probability a hypnozoite has died at time $t$ after inoculation, given by Equation (\ref{pdeath}).\\

\begin{figure}[ht]
\centering
\includegraphics[width=0.8\textwidth]{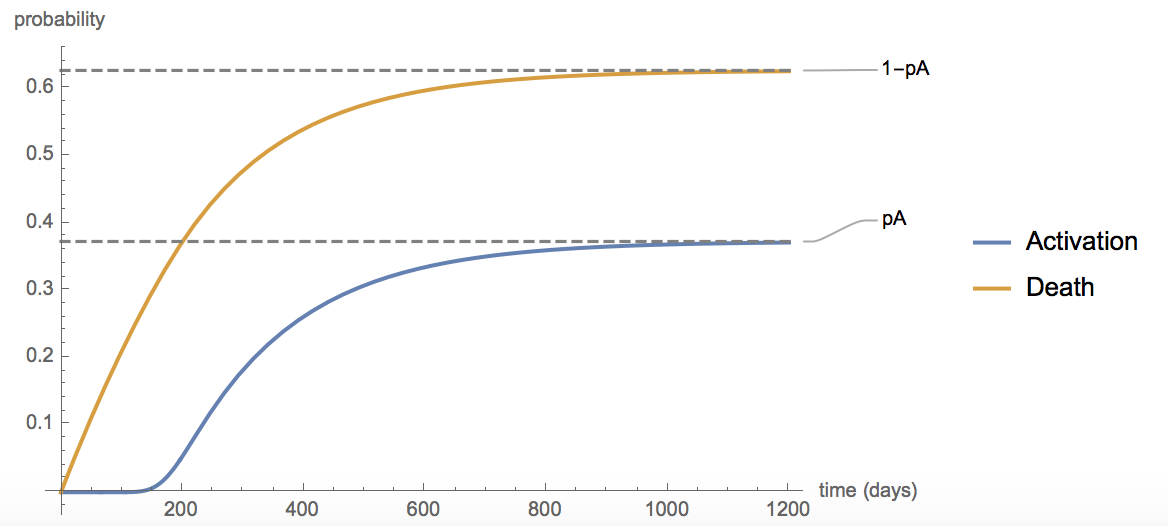}
\caption{Probabilities of activation and of death by time $t$ for a single hypnozoite given $\delta = 1/5 \text{ day}^{-1}, \mu = 1/442 \text{ day}^{-1}, \alpha = 1/325 \text{ day}^{-1}, k = 35$. For this choice of parameters, $p_{\text{active}}(t)$ and $p_{\text{death}}(t)$ {{ approach}} 0.373 and 0.627, respectively, as $t \to \infty$.}
\label{act_prob}
\end{figure}

Figure \ref{act_prob} illustrates the probability of activation, $p_{\text{active}}(t)$, and death, $p_{\text{death}}(t)$, over time for a single hypnozoite, using the same parameters as \textcite{whitemodel}. In the limit as $t \to \infty$, the probabilities of activation and death asymptotically approach $p_A$ (Equation (\ref{p_A})) and $1-p_A$, respectively.\\

In Section 4, we will use some of the results from this section to derive the dynamics when multiple hypnozoites are initially established by a single mosquito bite.

\section{Modelling hypnozoite dynamics for a single mosquito bite}\label{multiple}
To extend our model for a single hypnozoite to examine  dynamics for a single mosquito bite, which can lead to the establishment of multiple hypnozoites, we examine latency, death and activation for a group of hypnozoites inoculated at the same time. We do not account for variation in the time taken for hypnozoites to be established in the liver, assuming instead that each hypnozoite commences latency the moment it is inoculated. However, by adding an extra compartment (or compartments) to the model, it would be a relatively simple matter to account for this establishment time. We focus, in particular, on distributions for the time to first relapse ({{that is, the}} first activation event) and the duration of hypnozoite carriage ({{that is,}} the time to clearance of all hypnozoites). We first do this for a fixed number of hypnozoites and then extend our analysis so that the number of hypnozoites follows a geometric distribution.

\subsection{Relaxing collective dormancy to allow for independent hypnozoite behaviour}\label{white_diff}

{{It is an assumption in the work of \textcite{whitemodel} that all hypnozoites established by a mosquito bite progress through successive compartments of the latency phase in lock-step. In this work, we relax White's condition of collective dormancy, and instead assume complete statistical independence between all hypnozoites. Our model for a single hypnozoite is equivalent to White's model for a mosquito bite establishing $n=1$ hypnozoites, however, there are differences in the case of a mosquito bite establishing $n>1$ hypnozoites, noting that $n>1$ is almost certainly the case biologically.}\\

{Our model and White's model differ in that they have different state spaces. In White's model, it is sufficient to describe the state of the Markov chain by the number of hypnozoites remaining in the liver, and the number of the compartment in which these remaining hypnozoites are located (that is, compartments $1 \ldots k$ of the latency phase, non-latency compartment $k+1$). Hypnozoites are cleared from the liver due to death or activation, however, White's model tracks only the cumulative number of cleared hypnozoites; activated and dead hypnozoites are not distinguished. The state space for our model, in which each hypnozoites moves independently, is broader. We monitor each hypnozoite, which can occupy a state $j \in \{ 1, \ldots, k, \text{nonlatency}, \text{death}, \text{activation} \}$. However, a more parsimonious state space, as we observe, simply records the number of hypnozoites in each compartment.}\\

{The transition rates of our model and White's model also differ. In both models, the latency phase is governed by fixed parameters $k$ and $\delta$, which represent the number of latency compartments and the rate of progression through successive latency compartments respectively. In White's model, $n$ hypnozoites in compartment $j$ of the latency phase will collectively progress to compartment $j+1$ at rate $\delta$. In our model, however, we allow each hypnozoite to progress independently through successive compartments of the latency phase at rate $\delta$.}\\
}

\subsection{Complete characterisation of the hypnozoite reservoir for a single mosquito bite}\label{infqueue}
Suppose that a fixed $n$ hypnozoites have been established through a mosquito bite at time $t=0$. We make the assumption that each hypnozoite established by a mosquito bite behaves independently. We can therefore conceptualise our compartmental model as a network of infinite server queues.\\

We define
\begin{align*}
    \mathbf{X}(t) = (X_1(t), X_2(t), \dots, X_n(t))
\end{align*}
to be the state of the system, where the components $X_i(t)$ are independent and identically-distributed (i.i.d.) with PMF given by Equations (\ref{pm}), (\ref{nonlatent}), (\ref{pactive}) and (\ref{pdeath}). In this description, $X_i(t)$ is the compartment in which hypnozoite $i$ resides at time $t$.

\vspace{2mm}

Any state ${\bf X}(t)$ can be mapped to a vector ${\bf n}(t)$ of length $k+3$ whose $j$th component $n_j(t)$ gives the number of hypnozoites residing in compartment $j$ (either one of the $k$ latent compartments or the nonlatent, active or death compartments) at time $t$.

\vspace{2mm}

The vector ${\bf n}(t)$ must satisfy 
\begin{align*}
    n_\text{nonlatent}(t) + n_\text{active}(t) + n_\text{death}(t) + \sum^{k}_{m=1} n_m(t) = n,
\end{align*}
and it follows from the work of \textcite{harrison1981note} that ${\bf n}(t)$ has the multinomial distribution given by
\begin{align}
    \Pr[n_m(t)=l_m(t) \hspace{1mm} & \forall \hspace{1mm} m \in \{1, \dots, k\}, \hspace{1mm} n_\text{nonlatent}(t)=h, \hspace{1mm} n_\text{active}(t)=a, \hspace{2mm} n_\text{death}(t)=d] \notag \\
    = & \frac{n!}{h! a! d! \prod^{k}_{m=1} l{_m}!} \big( p_\text{nonlatent}(t) \big)^h  \big( p_\text{active}(t) \big)^a \big( p_\text{death}(t) \big)^d \prod^{k}_{m=1} \big( p_m(t) \big)^{l_m}.
\end{align}

\vspace{2mm}

For some subset $s \subset \lbrace1,2,\ldots,k,  \text{nonlatent, active, death}\rbrace$ let $Y_s(t)$ denote the total number of hypnozoites in the set $s$, that is $Y_s(t) = \sum_{j\in s} n_j(t)$. Let $p_s(t)$ denote the probability of a given hypnozoite being in any of the states in $s$. Then we note that $Y_s(t)$ has the binomial distribution, with PMF
\begin{align}
    \Pr[Y_s=m] = \genfrac(){0pt}{0}{n}{m} \big( p_s(t) \big)^m  \big(1 - p_s(t))^{n-m}.\label{binom}
\end{align}

\subsection{First relapse arising from a single mosquito bite} \label{first_relapse}
From Equation (\ref{binom}), the probability $P_{j}(t)$ that precisely $j$ hypnozoites will have activated by time $t$ is given by the Binomial distribution
\begin{align}
P_{j}(t) = \genfrac(){0pt}{0}{n}{j} p_{\text{active}}(t)^j [1-p_{\text{active}}(t)]^{n-j},
\end{align}
where $p_{\text{active}}(t)$ is given in Equation (\ref{pactive}). In the same way, the probability that exactly $j$ of the $n$ inoculated hypnozoites activate ({{that is,}} give rise to a relapse) in the limit as $t \to \infty$ is given by the Binomial distribution
\begin{align}
    \text{Pr($j$ relapses)} = \genfrac(){0pt}{0}{n}{j} (p_A)^j (1 - p_A)^{n-j}. \label{prob_j}
\end{align}
Therefore, the probability of at least 1 relapse as $t\to \infty$ is given by
\begin{align}
    \text{Pr($\geq$ 1 relapses)} = 1 - (1-p_A)^n, \label{more_than_one_relapse}
\end{align} 
and the expected number of relapses arising from an infected bite establishing $n$ hypnozoites is
\begin{align}
    n p_A = n \frac{\alpha}{\alpha+\mu} \Big(\frac{\delta}{\delta+\mu}\Big)^{k}, \label{npA}
\end{align}
where $p_A$ is the probability of activation before death for each hypnozoite, given by Equation (\ref{p_A}).\\

Now, suppose we are given that exactly $j$ relapses occur. Since hypnozoite activation times are i.i.d., the CDF for the  $i^{th}$ relapse is given by the probability of at least $i$ successes in a Bernoulli Trial with success probability $A(t)$:
\begin{align}
    P_{R_{i} | j}(t) = \sum^{j}_{m=i} \genfrac(){0pt}{0}{j}{m} A(t)^m (1-A(t))^{j-m}.
\end{align}
Recall that $A(t)$ is the probability that activation has occurred by time $t$, given activation occurs (Equation (\ref{A})).\\

In particular, given exactly $j$ relapses occur, the CDF for the first relapse is
\begin{align}
    P_{R_{1} | j}(t) = 1 - (1-A(t))^j. \label{r1j}
\end{align}

The probability that the first relapse has occurred by time $t$ is therefore
\begin{align}
    G_n(t) & = \Pr[1^{\text{st}} \text{ relapse by $t$}] \notag\\
            & = \sum^n_{j=1} \overbrace{\genfrac(){0pt}{0}{n}{j} (p_A)^j (1 - p_A)^{n-j}}^{\Pr[\text{$j$ relapses}]} \times \overbrace{[1 - (1 - A(t))^j] \vphantom{\genfrac(){0pt}{0}{n}{j}}}^{\Pr[1^{\text{st}} \text{ relapse by $t$} | \text{$j$ relapses}]} \notag\\
            & = [1 - (1 - p_A A(t))^n], \label{Gn}
\end{align}
where we have used Equations (\ref{prob_j}) and (\ref{r1j}). \\

Given {{ that}} at least one relapse occurs, we can use the above results to derive the CDF for the first relapse:
\begin{align}
     F_n(t) & = \Pr[1^{\text{st}} \text{ relapse by $t$}|\geq 1 \text{ relapses}] \notag\\
     & = \underbrace{\frac{1}{1-(1-p_A)^n}}_{1/\Pr[\geq 1 \text{ relapses}]} \underbrace{[1 - (1 - p_A A(t))^n] \vphantom{\frac{1}{1-(1-p_A)^n}}}_{\Pr[1^{\text{st}} \text{ relapse by $t$}]}, \label{Fn}
\end{align}
where we have used Equations (\ref{more_than_one_relapse}) and (\ref{Gn}). \\

By definition, given at least one relapse occurs, the expected time to first relapse for a bite establishing $n$ hypnozoites is given by
\begin{align}
    T^{n}_{r} = \int^{\infty}_{0} \Big( 1 - F_n(t) \Big) dt. \label{Trn}
\end{align}

\begin{figure}[!ht]
\centering
\includegraphics[width=0.85\textwidth]{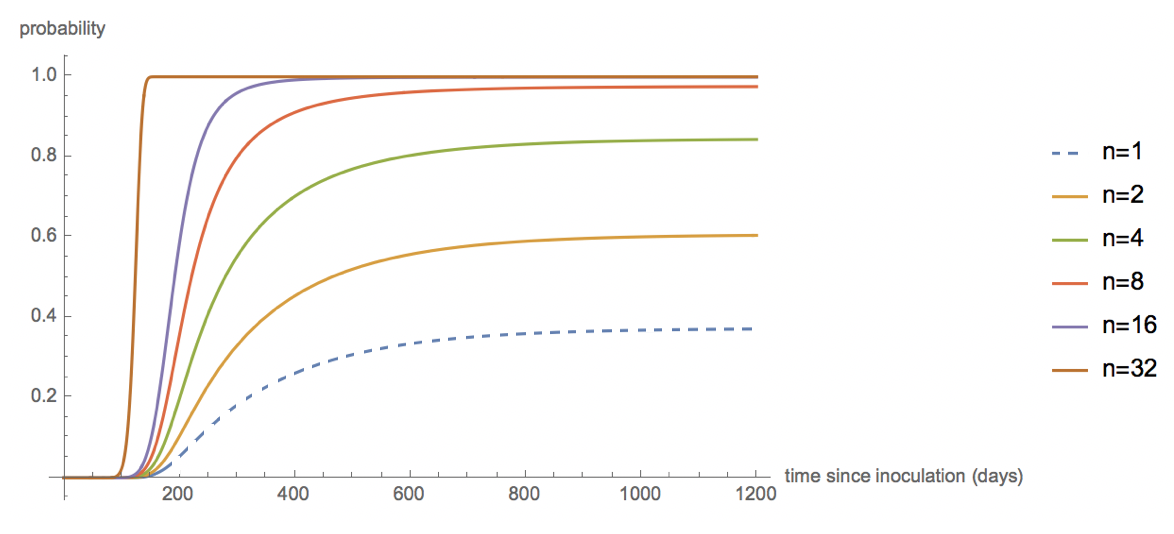}
\caption{Distribution functions for the time to first relapse, $G_n(t)$ (Equation (\ref{Gn})), for bites establishing $N$=1, 2, 4, 8, 16, and 32  hypnozoites. Parameter values used are $\delta = 1/5 \text{ day}^{-1}, \mu = 1/442 \text{ day}^{-1}, \alpha = 1/325 \text{ day}^{-1}, k=35$.}\label{first_relapse_diff_prob}
\end{figure}

Figure \ref{first_relapse_diff_prob} illustrates the probability $G_n(t)$ that the first relapse has occurred by time $t$ for mosquito bites establishing a different number of hypnozoites, $n$. In the limit $t \to \infty$, the probability that the first relapse has occurred asymptotically approaches $(1 - (1-p_A)^n)$ (Equation (\ref{more_than_one_relapse})). A relapse is not guaranteed to arise from each mosquito bite, particularly for mosquito bites establishing only a few hypnozoites. As the number of inoculated hypnozoites increases, the probability of experiencing at least one relapse approaches one. In the case $n=1$ (dotted blue curve), we recover the distribution for the activation of a single hypnozoite that is illustrated in Figure \ref{act_prob}.\\

\begin{figure}[!ht]
\centering
\includegraphics[width=0.75\textwidth]{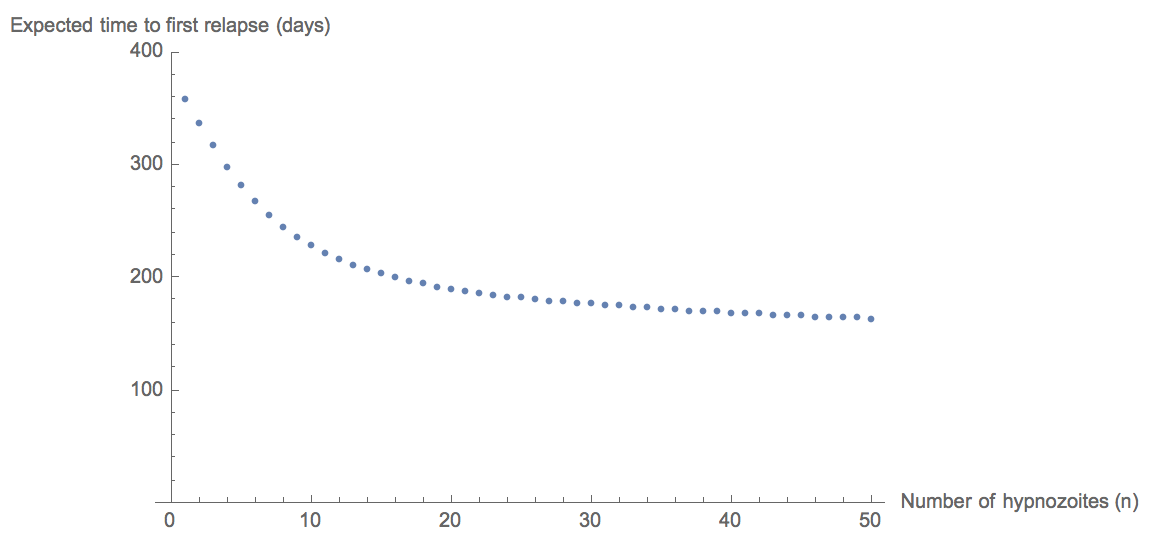}
\caption{Expected time to first relapse $T^{n}_{r}$ (Equation (\ref{Trn})), given at least 1 relapse occurs, as a function of the number of hypnozoites $n$ established by a mosquito bite. Parameter values used are $\delta = 1/5 \text{ day}^{-1}, \mu = 1/442 \text{ day}^{-1}, \alpha = 1/325 \text{ day}^{-1}, k = 35$.} \label{expected_time_first_relapse}
\end{figure}

Figure \ref{expected_time_first_relapse} illustrates the expected time to first relapse given at least one relapse occurs, $T^{n}_r$, as a function of the size of the hypnozoite inoculum. The expected time to first relapses decreases monotonically as the size of the hypnozoite inoculum increases, with additional hypnozoites having progressively smaller effects on the time to first relapse.\\

\subsection{Duration of hypnozoite carriage for a single mosquito bite}
We now investigate the duration of hypnozoite carriage for a single infected mosquito bite, a biologically-relevant quantity; that is, we derive an expression for the time to clearance of all hypnozoites.  The probability that a hypnozoite has been cleared at time $t$ after inoculation is given by
\begin{align*}
    p_{\text{clear}}(t) = p_{\text{active}}(t) + p_{\text{death}}(t)
\end{align*}
since death and activation are the two absorbing states. The form of $p_{\text{active}}(t)$ and $p_{\text{death}}(t)$ are given by Equations (\ref{pactive}) and (\ref{pdeath}), respectively.\\

Since hypnozoite clearance times are i.i.d., as in Section \ref{first_relapse}, the distribution function for the clearance of the $i^{th}$ hypnozoite is given by the probability of at least $i$ successes in a Bernoulli trial with success probability $p_\text{clear}(t)$:
\begin{align}
    C_{i | n}(t) = \sum^{n}_{m=i} \genfrac(){0pt}{0}{n}{m} (p_\text{clear}(t))^m (1 - p_\text{clear}(t))^{n-m}.
\end{align}

Therefore, the CDF for total hypnozoite clearance $C_n(t)$ for a mosquito bite establishing $n$ hypnozoites is given by
\begin{align}
    C_n(t) = p_{\text{clear}}(t)^n, \label{C_n}
\end{align}
with expected time to clearance
\begin{align}
    T^{n}_c = \int^{\infty}_{0} \Big( 1 - p_{\text{clear}}(t)^n \Big) dt.  \label{Tn_c}
\end{align}

Let $L(t)$ denote the number of hypnozoites remaining in the liver ({{that is,}} not yet cleared) by time $t$. From the Binomial distribution, we expect $n p_{\text{clear}}(t)$ hypnozoites to be cleared by time $t$. Thus
\begin{align}
    \mathbb{E}[L(t)] = n(1 - p_{\text{clear}}(t))
\end{align}
hypnozoites are expected to remain in the liver ({{that is}}, contribute to the hypnozoite reservoir) at time $t$.\\

\begin{figure}[!ht]
\centering
\includegraphics[width=0.85\textwidth]{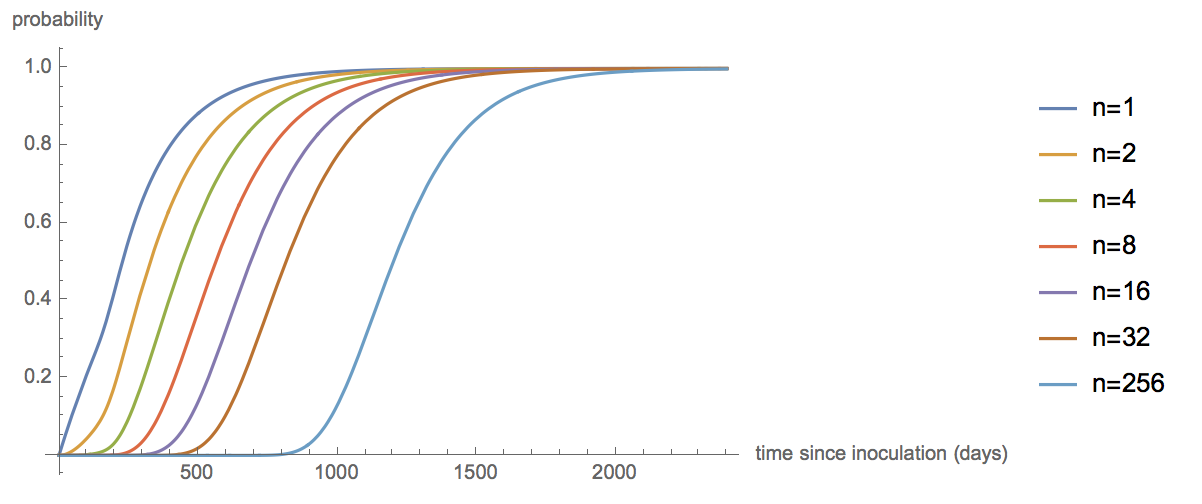}
\caption{CDFs for the clearance time, $C_n(t)$, for bites establishing $N$=1, 2, 4, 8, 16, 32 and 256 hypnozoites (Equation (\ref{C_n})). Parameters used are $\delta = 1/5 \text{ day}^{-1}, \mu = 1/442 \text{ day}^{-1}, \alpha = 1/325 \text{ day}^{-1}, k = 35$.}\label{clearance_cdf}
\end{figure}

\begin{figure}[!ht]
\centering
\includegraphics[width=0.85\textwidth]{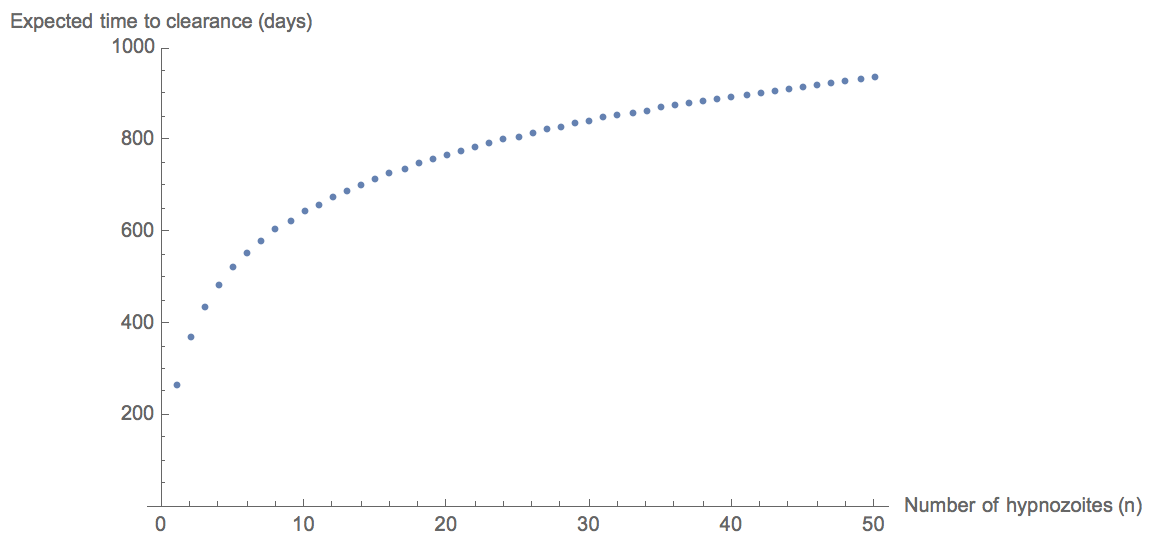}
\caption{Expected clearance time, $T^{n}_{c}$, as a function of the number of hypnozoites established, $n$, by a mosquito bite (Equation (\ref{Tn_c})). Parameters used are $\delta = 1/5 \text{ day}^{-1}, \mu = 1/442 \text{ day}^{-1}, \alpha = 1/325 \text{ day}^{-1}, k = 35.$}\label{expected_time_clearance}
\end{figure}

Figure \ref{clearance_cdf} illustrates distributions for the clearance times, $C_n(t)$, for mosquito bites establishing different numbers of hypnozoites (Equation (\ref{C_n})) and shows that as $t \to \infty$, the hypnozoites are cleared, but that the probability of clearance at any time is higher for fewer initial hypnozoites. Figure \ref{expected_time_clearance} illustrates the expected time to total clearance, $T^{n}_c$, as a function of the hypnozoite inoculum (Equation (\ref{Tn_c})). The duration of hypnozoite carriage increases monotonically with the inoculum size. We see that additional hypnozoites have progressively smaller effects on the duration of hypnozoite carriage. \\

\subsection{Accounting for variation in the number of hypnozoites per mosquito bite}
We now relax the assumption of fixed $n$ and allow the number of hypnozoites established per bite to be geometrically-distributed with mean $N$, similarly to \textcite{whitemodel}. The probability of $n$ hypnozoites being established by a bite is thus
\begin{align}
    \Pr[\text{$n$ hypnozoites}] = \frac{1}{N+1} \Big(\frac{N}{N+1}\Big)^n. \label{prob_n}
\end{align}\

By the conditional expectation formula, the expected number of relapses in the limit as $t \to \infty$ arising from an infected bite is therefore
\begin{align}
    R = \mathbb{E}[\mathbb{E}[\text{relapses} | \text{$n$ hypnozoites}]] = \mathbb{E}[n p_A] = N p_A
\end{align}
which is the analogue of Equation (\ref{npA}).

The probability of at least one relapse arising from any bite (analogous to Equations (\ref{more_than_one_relapse})) is given by
\begin{align}
    \text{Pr($\geq$ 1 relapses)} = \sum^{\infty}_{n=0} \overbrace{\frac{1}{N+1} \Big(\frac{N}{N+1}\Big)^n}^{\Pr[\text{$n$ hypnozoites}]} \overbrace{\Big(1 - (1-p_A)^n \Big)}^{\Pr[\geq 1 \text{ relapse} | \text{$n$ hypnozoites}]} = \frac{N p_A}{1 + N p_A}, \label{prob_more_than_1_relapse}
\end{align}
where the RHS has been simplified by noting that the sum comprises of two geometric series.

Hence, given at least one relapse occurs, the expected time to first relapse is given by
\begin{align}
    T_r^N & = \overbrace{\frac{1 + N p_A}{N p_A}}^{1/\Pr[\geq 1 \text{ relapses}]} \sum^{\infty}_{n=1} \overbrace{\frac{1}{N+1} \Big(\frac{N}{N+1}\Big)^n}^{\Pr[\text{$n$ hypnozoites}]} \times \overbrace{\int^{\infty}_{0} \Big[\frac{(1-p_A A(t))^n - (1-p_A)^n}{1-(1-p_A)^n} \Big ] dt,}^{\mathbb{E} [\text{time to first relapse} | \text{$n$ hypnozoites}]}
\end{align}
where we have used Equations (\ref{Trn}), (\ref{prob_n}) and (\ref{prob_more_than_1_relapse}).\\

The mean duration of hypnozoite carriage is then
\begin{align}
    T_c^N & = \sum^{\infty}_{n=1} \overbrace{\frac{1}{N+1} \Big(\frac{N}{N+1}\Big)^n}^{\Pr[\text{$n$ hypnozoites}]} \times \overbrace{\int^{\infty}_{0} \Big[ 1 -  p_{\text{clear}}(t)^n \Big ] dt}^{\mathbb{E} [\text{clearance time} | \text{$n$ hypnozoites}]} \notag \\ 
    & = \int^{\infty}_{0} \Big[ 1 - 
    \frac{1}{1 + N (1 - p_{\text{clear}(t)})} \Big] dt,
\end{align}
where we have used Equations (\ref{prob_n}) and (\ref{Tn_c}), and the RHS has been simplified similarly to Equation (\ref{prob_more_than_1_relapse}) after interchanging the summation and the integral.

\subsection{Simulating an exemplar patient}
\begin{figure}[ht]
\centering
\includegraphics[width=0.80\textwidth]{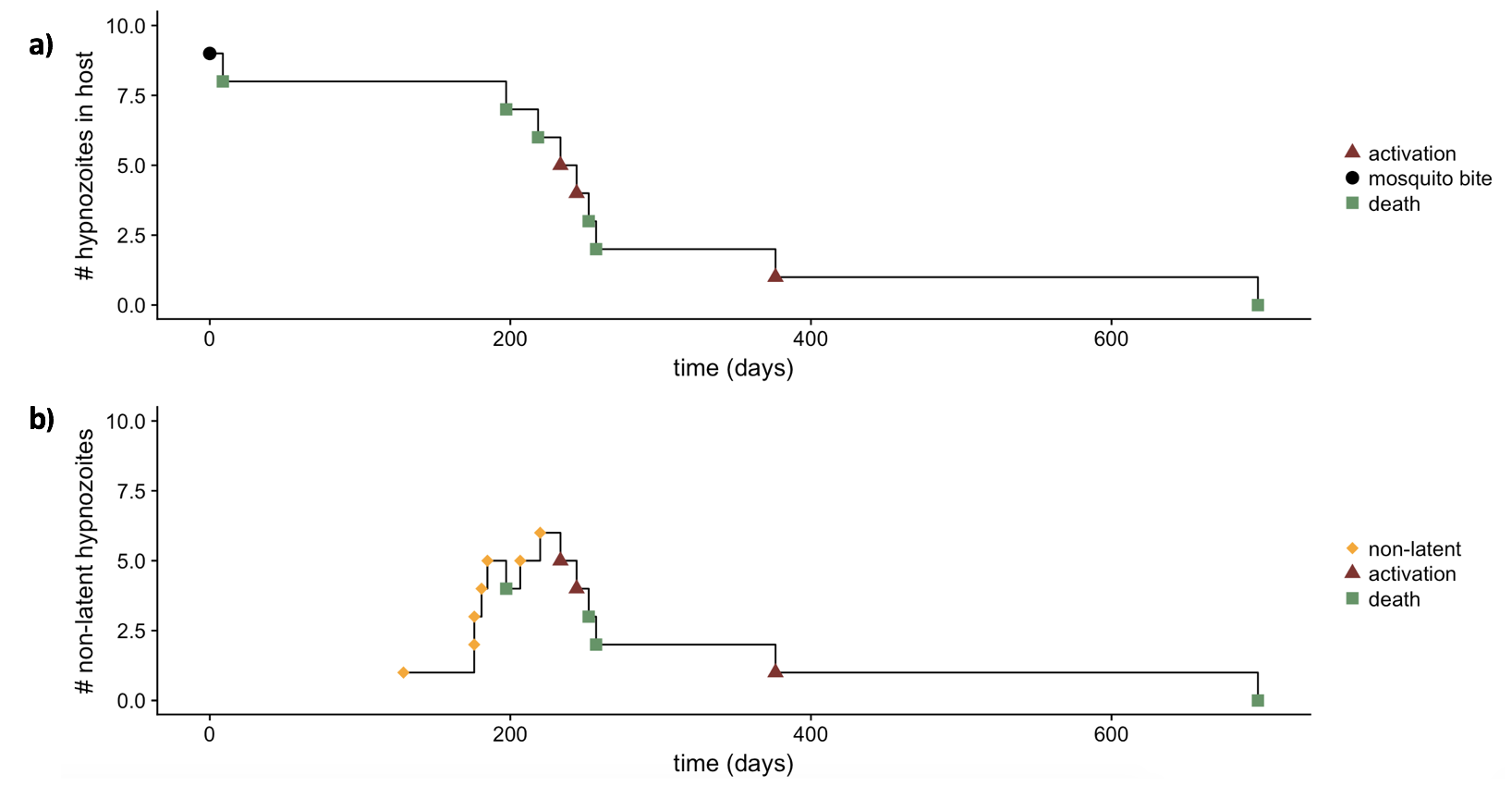}
\caption{Simulated sample path for a single mosquito bite establishing 9 hypnozoites. Parameters used are $\delta = 1/5 \text{ day}^{-1}, \mu = 1/442 \text{ day}^{-1}, \alpha = 1/325 \text{ day}^{-1}, k = 35$.}\label{sample_path}
\end{figure}

We now illustrate the dynamics of hypnozoite activation and clearance for an exemplar patient. Figure \ref{sample_path} shows a typical sample path for the model presented in Section \ref{single} using direct stochastic simulation (as in the Doob-Gillespie algorithm). At time $t=0$, an individual is inoculated with nine hypnozoites (black circle). Figure 7(a) tracks the number of hypnozoites remaining in the human host, with hypnozoite clearance achieved after approximately two years (consistent with Figure \ref{expected_time_clearance} for nine hypnozoites). Six hypnozoites die without activating (green squares), while the remaining three hypnozoites activate (brown triangles), leading to the initiation of three blood-stage infections. Figure 7(b) tracks the number of nonlatent hypnozoites in the human host ({{that is,}} compartment $k+1$) and shows that only seven hypnozoites become nonlatent (yellow circles). The remaining two hypnozoites die during latency. Hypnozoites become susceptible to activation only after the latency phase is complete; however, four of the seven hypnozoites that do complete latency die before they can activate. The first activation event occurs just after 200 days, consistent with Figure \ref{expected_time_first_relapse} for nine hypnozoites.

\vspace{2mm}

Since the White model enforces a collective latency period, with all hypnozoites established by a mosquito bite emerging from dormancy at the same instant, a sample path for their model would show a steep jump in the number of nonlatent hypnozoites at the end of the collective dormancy period ({{that is,}} all yellow dots would coincide in Figure 7(b)). Thereafter, nonlatent hypnozoites would be cleared progressively due to either death or activation which is modelled independently for each hypnozoite.

\section{A numerical comparison of nonlatent hypnozoite distributions}\label{comparison}
To compare our model to White's model for long-latency strains (see Equation (7) in \cite{whitemodel}), we consider the respective distributions of nonlatent hypnozoites in the liver ({that is,} the number of hypnozoites susceptible to activation). In the case that a single hypnozoite is established by a mosquito bite, our model is equivalent to the White model. However, an infected mosquito bite is expected to establish multiple hypnozoites, and so in this case there are important differences in the dynamics of the two models. \\

To aid in our comparison, we have derived an analytic solution to the White model by noting that movement through the compartments of the latency phase is independent {{ from}} the number of hypnozoites remaining in the liver (details in the Appendix). In short, the solution to White's model relies on the observation that progress through compartments of the latency phase is governed by a Poisson process, while an independent pure death process governs the number of hypn{o}zoites remaining in the liver during latency. \\

Suppose $N$ hypnozoites are established by a mosquito bite at time $t=0$. For our model, the probability $H^{\text{new}}_{j}(t)$ that $j$ nonlatent hypnozoites remain in the reservoir at time $t$ after inoculation is given by the Binomial distribution
\begin{align}
H^{\text{new}}_{j}(t) = \genfrac(){0pt}{0}{N}{j} p_{\text{nonlatent}}(t)^j [1-p_{\text{nonlatent}}(t)]^{N-j}, \label{jnonlat}
\end{align}
as per Equation (\ref{binom}), where $p_{\text{nonlatent}}(t)$ is the probability of a hypnozoite being nonlatent at time $t$ after inoculation, given by Equation (\ref{nonlatent}).

The expected number of nonlatent hypnozoites in the reservoir at time $t$ is then
\begin{align}
\mathbb{E}[H^{\text{new}}(t)] = N p_{\text{nonlatent}}(t).
\end{align}

For a single bite establishing $N$ hypnozoites, given the same parameters $\delta, \alpha, \mu$ and $k$, it can be shown (see the Appendix) that the expected number of nonlatent hypnozoites at a given time is identical for our model and White's model
\begin{align*}
   \mathbb{E}[H^{\text{new}}(t)] = \mathbb{E}[H^{\text{White}}(t)] = N p_{\text{nonlatent}(t)}.\
\end{align*}

\vspace{2mm}

In the steady state, zero nonlatent hypnozoites remain in the liver as death and activation are absorbing states in both models. However, there are important differences in the transient distributions of nonlatent hypnozoites.

\vspace{2mm}

In the early stages of hypnozoite carriage, our model predicts a higher probability of a small number of hypnozoites being nonlatent since we account for the possibility of a subset of hypnozoites emerging relatively quickly from dormancy. However, noting that the expected number of nonlatent hypnozites is identical for both models, we see that the White model is weighted towards a higher number of concurrently nonlatent hypnozoites in the early stages of hypnozoite carriage: while there is a lower chance that hypnozoites will have emerged from dormancy, the probability of hypnozoite death is also lower. Hence, in White's model, we would expect the number of hypnozoites emerging from dormancy in lock-step to increase as the duration of dormancy increases.

\vspace{2mm}

\begin{figure}[ht]
\centering
\includegraphics[width=\textwidth]{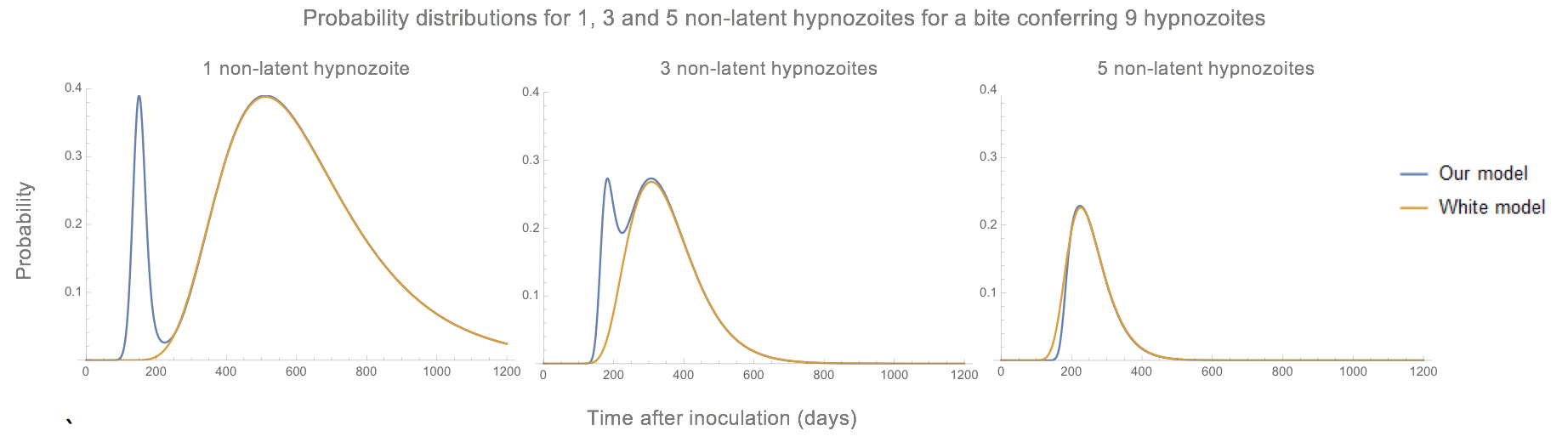}
\caption{Comparisons of the probability distributions for 1, 3 and 5 nonlatent hypnozoites, given an initial inoculation of 9 hypnozoites and parameters $\delta = 1/5 \text{ day}^{-1}, \mu = 1/442 \text{ day}^{-1}, \alpha = 1/325 \text{ day}^{-1}, k = 35$. Probability distributions for our model are shown in blue, while those for White's model are shown in yellow.}
\label{white_nl_comp}
\end{figure}

Given an initial inoculation of 9 hypnozoites, the probability distributions for the number of nonlatent hypnozoites time $t$ after inoculation predicted by our model (Equation (\ref{jnonlat})) and White's model converge after 200 days for the parameters used by \textcite{whitemodel}, as shown in Figure \ref{white_nl_comp}. Prior to this period, our model predicts a significantly higher probability of few ({{that is,}} 3 or less) nonlatent hypnozoites, while White's model is weighted towards a higher number of concurrently nonlatent hypnozoites.

\begin{figure}[ht]
\centering
\includegraphics[width=0.95\textwidth]{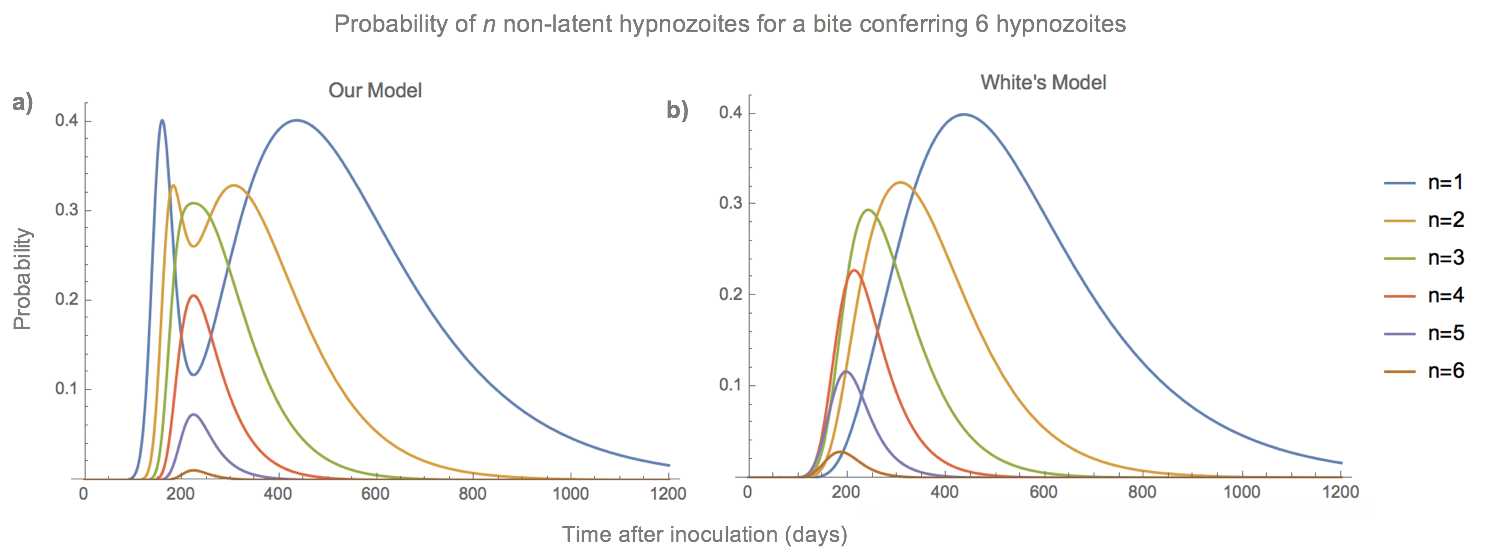}
\caption{Probability of $n$ nonlatent hypnozoites for a bite establishing 6 hypnozoites, given parameters $\delta = 1/5 \text{ day}^{-1}, \mu = 1/442 \text{ day}^{-1}, \alpha = 1/325 \text{ day}^{-1}, k = 35$ for our model (a) and White's model (b).}
\label{bite_6_hyp}
\end{figure}

Our probability distributions for small numbers of nonlatent hypnozoites also exhibit bimodal behaviour that is absent in White's model. Progression through the latency phase in lock-step precludes bimodality in White's model: if the duration of the dormancy period is short, we would expect more hypnozoites to be concurrently nonlatent since the probability of hypnozoites being cleared due to death is lower. Hence, the probability distribution for $n$ nonlatent hypnozoites peaks earlier, but with a smaller amplitude, than the probability distribution for $(n-1)$ nonlatent hypnozoites, as shown in Figure \ref{bite_6_hyp}(b).

\vspace{2mm}

Bimodality for small numbers of nonlatent hypnozoites in our model, in contrast, arises from the independent progression of hypnozoites through the latency phase. The probability of several nonlatent hypnozoites first peaks in the earlier stages of hypnozoite carriage because of the possibility a subset of hypnozoites have undergone relatively short dormancy periods, as shown in Figure \ref{bite_6_hyp}(a). As time progresses, the probability of hypnozoite clearance increases, leading to a decreasing probability of a large number of hypnozoites being present in the liver. Therefore, in the later stages of hypnozoite carriage, another peak is seen in the probability that a small number of hypnozoites are concurrently nonlatent.

\vspace{2mm}

\begin{figure}[ht]
\centering
\includegraphics[width=0.95\textwidth]{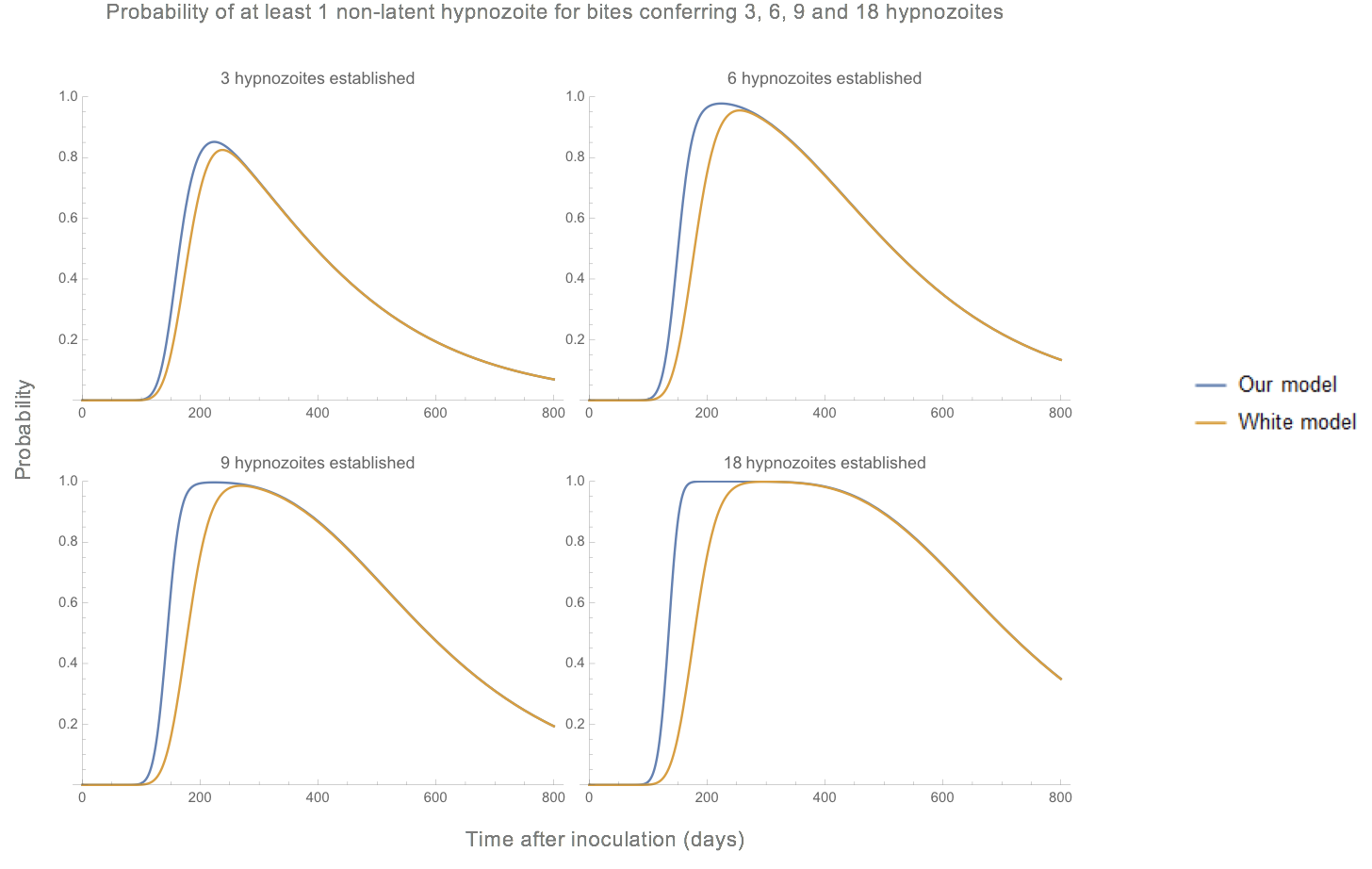}
\caption{Probability of at least one nonlatent hypnozoite for different hypnozoite inoculations, given parameters $\delta = 1/5 \text{ day}^{-1}, \mu = 1/442 \text{ day}^{-1}, \alpha = 1/325 \text{ day}^{-1}, k = 35$ for White's model (yellow) and our model (blue).}
\label{min_1_nonlat}
\end{figure}

The differing dynamics of our model and White's model have biological implications. For instance, the probability of at least one nonlatent hypnozoite, which serves as a metric for the risk of relapsing infection, peaks earlier and remains elevated for a significantly longer period in our model, as shown in Figure \ref{min_1_nonlat}. As the number of hypnozoites established through a mosquito bite increases, the differences between the models are amplified.

\vspace{2mm}

We would also expect that hypnozoite activation events are more clustered in White's model, since hypnozoites established through the same mosquito bite become susceptible to activation at the same instant. Our model, in contrast, allows for more spacing between relapses since we account for variation in the dormancy phase between hypnozoites established through the same bite.

\section{Discussion}\label{discussion}
The dynamics of the hypnozoite activation underpin the epidemiology of \textit{Plasmodium vivax}, with relapses constituting a significant proportion of the infection burden \parencite{howes2016global}. Although hypnozoite activation is known to trigger relapses, underlying biological processes driving hypnozoite activation are poorly understood. Here, we have developed a within-host model of hypnozoite activation adapted to long-latency strains of \textit{Plasmodium vivax}. We have modelled activation-clearance dynamics for a single hypnozoite through a continuous-time Markov chain, conceptualising latency as a progression through successive compartments, with absorbing states death and activation. To model the contribution of a single mosquito bite to the hypnozoite reservoir and possible relapses, we have examined activation-clearance dynamics after a mosquito bite, considering both fixed and variable numbers of hypnozoites. We have derived analytic distributions for the time to first relapse and the duration of hypnozoite carriage for a given mosquito bite, both of which are quantities of epidemiological and biological relevance.\\

Our work has extended that of \textcite{whitemodel}, who developed a model of hypnozoite dynamics for long-latency strains. By removing the requirement for hypnozoites established through the same mosquito bite to progress through compartments of the latency phase in lock-step, we have captured a more biologically-reasonable framework, whereby hypnozoites inoculated at the same time can emerge from dormancy at different time points. Our model allows dynamics of both hypnozoite activation and clearance to be monitored. The White model, however, principally monitors the number of hypnozoites in any given compartment of the latency phase. Extending the White model to track the number of hypnozoite activations arising from infected mosquito bites introduces further complexity, potentially limiting the tractability of the White model in more complex frameworks. Hypnozote activation events, in contrast, can be tracked readily using our framework. 

\vspace{2mm}

Our model examines a baseline scenario in which hypnozoite activation is assumed to occur at a constant rate post-latency. Hypothesised triggers for hypnozoite activation, including systemic febrile illness \parencite{whiteimwong} and bites from potential mosquito vectors \parencite{hulden} have not been considered. We have assumed, moreover, that each hypnozoite activation gives rise to a relapse. We have not accounted for factors like immunity that may prevent an activated hypnozoite from giving rise to a clinical or detectable blood infection \parencite{whiteimwong}.

\vspace{2mm}

Our model, nonetheless, provides a basis for exploring broader dynamics of hypnozoite clearance and activation. Our model can be readily embedded in epidemiological frameworks and  multi-scale models to examine processes like immunity and transmission. Analytic solutions for hypnozoite activation and clearance at the within-host scale may enhance the tractability of statistical inference, which can be highly computationally intensive for numerical models \parencite{garira2018primer}. Statistical inference for models on an epidemiological scale, in particular, is likely to be more feasible with an underlying analytic framework.

\vspace{2mm}

\section*{Acknowledgements}
S. Mehra acknowledges funding from the Australian Mathematical Sciences Institute (AMSI) Vacation Research Scholarships 2018/2019.  J.M. McCaw's research is supported by the Australian Research Council (ARC) Discovery Project DP170103076.  J.A. Flegg's research is supported by the ARC DECRA Fellowship DE160100227.  P.G. Taylor's research is supported by the ARC Laureate Fellowship FL130100039 and the ARC Centre of Excellence for the Mathematical and Statistical Frontiers (ACEMS).

\printbibliography
\end{document}


\setlength{\abovedisplayskip}{4pt}
\setlength{\belowdisplayskip}{10pt}
\setlength{\abovedisplayshortskip}{4pt}
\setlength{\belowdisplayshortskip}{10pt}

\section{ODEs governing the dynamics of a single hypnozoite}
The expansion of the Kolomogrov Forward Differential Equation (see Equation (3) in the main text) yields the following system of ODEs governing the dynamics of a single hypnozoite:
\begin{align*}
\frac{dp_{1}}{dt} &= {} -(\mu+\delta)p_{1}(t)\\
\frac{dp_{m}}{dt} &= {} -(\mu+\delta)p_{m}(t) + \delta p_{m-1}(t), \ m \in [2, k]\\
\frac{dp_{\text{nonlatent}}}{dt} &= {} -(\alpha+\mu)p_{\text{nonlatent}}(t) + \delta p_{k}(t) \\
\frac{dp_{\text{active}}}{dt} &= {} \alpha p_{\text{nonlatent}}(t)\\
\frac{dp_{\text{death}}}{dt} &= {} \mu \sum_{i=1}^{k} p_{i}(t) + \mu p_{\text{nonlatent}}(t). 
\end{align*}

\section{Analytic Solution to White's Model}
Here, we present an analytic solution to the within-host model of temperate relapses (i.e. long-latency strains of hypnozoites) developed by \textcite{whitemodel} (see Equation (7) in their paper). Hypnozoites are assumed to undergo a latency phase before they are capable of activation. The latency phase is modelled through a series of $k$ compartments, with hypnozoites progressing through successive compartments at rate $\delta$. Hypnozoites conferred through the same mosquito bite progress through compartments of the latency phase in lock-step, but die independently at rate $\mu$. In compartment $k$ of the latency phase, individual hypnozoites may either die or activate independently at rates $\mu$ and $\alpha$ respectively. \\

Suppose $N$ hypnozoites are conferred through a mosquito bite at time zero. Let $L^j_i(t)$ denote the probability that there are $i$ hypnozoites in compartment $j$ of the latency phase at time $t$, and $H_i$ denote the probability that $i$ hypnozoites are in the relapsing phase at time $t$. Our notation is identical to that of \textcite{whitemodel}, except here we distinguish
\begin{equation}
    H^{N}_{0} = H_0 + \sum^{k}_{j=1} L^j_0.
\end{equation}

A schematic of their model is shown in Figure 1. 

\begin{figure}[ht]
\centering
\includegraphics[width=0.7\textwidth]{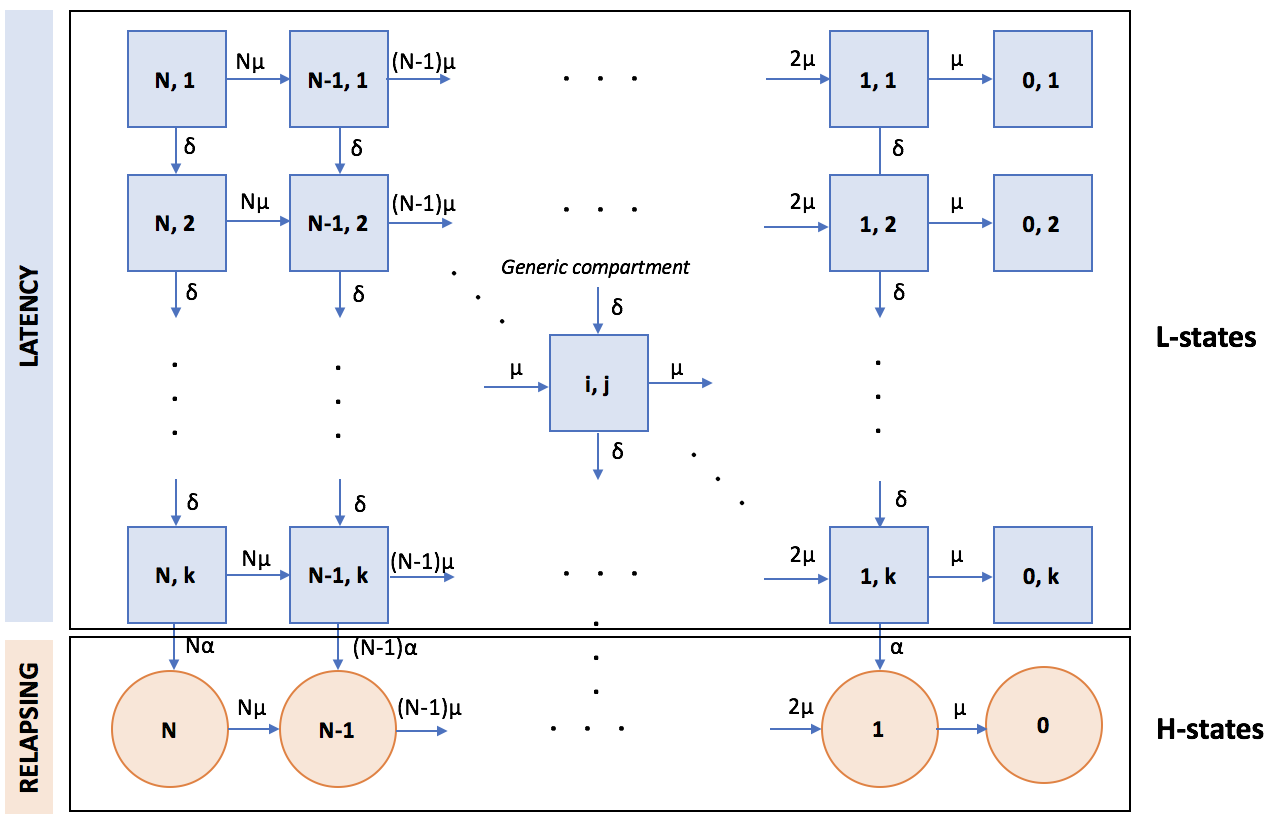}
\caption{Schematic for model of dynamics for $N$ hypnozoites developed by \textcite{whitemodel}. Hypnozoites conferred through the same mosquito bite progress through $k$ compartments of the latency phase in lock-step at rate $\delta$ before entering a relapsing phase. Individual hypnozoites die independently at constant rate $\mu$, and can activate independently at rate $\alpha$ from compartment $k$ of the latency phase.}
\label{hypnozoite_activation}
\end{figure}

\subsection{Derivation of State Probabilities}

The Master Equations governing this process are
\begin{align}
   &  \frac{d L^j_i}{dt} = \delta \big( L^{j-1}_i - L^{j}_{i} \big) + \mu \big( (i+1) L^j_{i+1} - i L^j_i) \text{ where } i \in \{ 0,...,N \} \text{, } j \in \{ 1, ..., k \},\\
   & \frac{dH_i}{dt} = \delta L^m_i + (\mu + \alpha) \big( (i+1) H_{i+1} - iH_i \big) \text{ where } i \in \{ 0,...,N \},
\end{align}
with initial condition
\begin{align}
    L^{1}_{N}(0) = 1 \text{, } L^{j}_{i} = H_i(0) = 0 \text{ for all other states.}
\end{align}

Due to the connectivity and rates of the model, we do not need boundary compartments with special rules. Instead, we extend the indices to $\infty$, and note that $L^j_i$ for $j<1$ and $i>N$ is 0, and $H_i$ for $i>N$ is also zero.\\

\textbf{Aside: Markovian Death Process}\\
We shall need the analytic solution to the Markovian death process in our solution. Consider a pure Markovian death process with death rate $\lambda$, as shown in Figure 2. 
\begin{figure}[ht]
\centering
\includegraphics[width=0.6\textwidth]{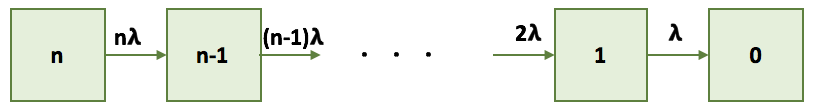}
\caption{Schematic for Markovian death process with death rate $\lambda$.}
\label{markovian_death}
\end{figure}

Suppose that the system is in state $n$ at time $t_0$. Then the probability of the system being in state $i \leq n$ at time $t>t_0$ is \parencite{epstein2008mathematical}
\begin{align}
    P(i, t | n, t_0) = K(i, t, n, t_0, \gamma) = \genfrac(){0pt}{0}{n}{i} \exp{(-n \gamma (t-t_0))} \big[ \exp{(\gamma (t-t_0))} - 1 \big]^{n-i}.
\end{align}

\vspace{3mm}

Returning to the model of the hypnozoite reservoir, we note that, while $H_i$ depends on $L^j_i$, $L^j_i$ can be solved independently from $H_i$. We therefore begin by solving Equation (2) for $L^j_i$. Examining the compartment structure (Figure 1), we deduce that the indices for $i$ and $j$ are independent, that is
\begin{align}
    L^j_i = L_i L^j,
\end{align}
where $L_i$ is the probability of being in an $L$-state $L^j_i$ for any $j$, and $L^j$ is the probability of being in an $L$-state $L^j_i$ for any $i$. We can verify this by substituting Equation (6) into Equation (2)
\begin{align}
    L_i \frac{dL^j}{dt} + L^j \frac{dL_i}{dt} = \delta L_i \big( L^{j-1} - L^j \big) + \mu L^j \big( (i+1) L_{i+1} - i L_i \big),
\end{align}
which yields
\begin{align}
    \frac{1}{L^j} \frac{dL^j}{dt} - \frac{\delta}{L^j} \Big( L^{j-1} - L_j) = \frac{1}{L_i} \frac{dL_i}{dt} - \frac{\mu}{L_i} \Big( (i+1) L_{i+1} - iL_i) = 0.
\end{align}
Since Equation (6) holds for all $i$ and $j$, the LHS and RHS are equal to a common constant which must be 0 so that probability is conserved. We thus obtain the independent processes
\begin{align}
    \frac{dL^j}{dt} = \delta \big(L^{j-1} - L^j \big) \text{ where } j \in \{ 1, ..., m \},\\
    \frac{dL_i}{dt} = \mu \big( (i+1)L_{i+1} - iL_i \big) \text{ where } i \in \{0, 1, ... \},
\end{align}
with initial conditions
\begin{align}
    & L^1(0) = 1 \text{ and } L^j(0)=0 \hspace{2mm} \forall j \neq 1,\\
    & L_N(0) = 1 \text{ and } L_i(0)=0 \hspace{2mm} \forall i \neq N.
\end{align}

\vspace{2mm}

Examining Equation (9), we see that $L^j$ is a simple queue which never gets shorter and increases in size at a constant rate $\delta$, i.e. it is a Poisson counting process. Hence
\begin{align}
    L^j(t) = \frac{(\delta t)^{j-1}}{(j-1)!} e^{-\delta t},
\end{align}
which is the PMF of a Poisson process with no $j=0$ component.\\

$L_i$, however, is a pure Markovian death process (see Equation (5)) with rate $\mu$ that is in state $N$ at time zero, and thus has solution
\begin{align}
    L_i(t) = K(i, t, N, 0, \mu) = \genfrac(){0pt}{0}{N}{i} \exp{(-N \mu t)} \big[ \exp{(- \mu t)} - 1 \big]^{N-i}.
\end{align}

Substituting Equations (13) and (14) into (6), we obtain the joint probability
\begin{align}
    L^j_i(t) = \genfrac(){0pt}{0}{N}{i} \frac{(\delta t)^{j-1}}{(j-1)!} \exp{(-(\delta + N \mu) t)} \big[ \exp{(- \mu t)} - 1 \big]^{N-i}.
\end{align}

\vspace{4mm}

We now shift our attention to Equation (3) for $H_i$. Since these equations are linear, we can use a Greens function approach, setting $G_i$ to the solution of
\begin{align}
    \frac{dG_i}{dt} = (\mu + \alpha) \big( (i+1) G_{i+1} - iG_i \big) + \delta_{i_0}(t-t_0), \text{ } G_i(0)=0, i \in \{ 0, ..., N \},
\end{align}
where $\delta_{i_0}$ is the Dirac delta function if $i=i_0$ and 0 otherwise. $H_i$ can be written in terms of $G_i(t, i_0, t_0)$
\begin{align}
    H_i(t) = \sum^{N}_{i_0=0} \int^{\infty}_{0} \delta L^k_{i_0}(t_0) G_i(t, i_0, t_0) dt_0.
\end{align}

Since $G_i(0)=0 \hspace{2mm} \forall i$, solving Equation (16) for $t<t_0$ we see that 
\begin{align}
    G_i(t, i_0, t_0) = 0 \text{ for } t<t_0.
\end{align}

Integrating $G$ from $t=t_0 - \epsilon$ to $t=t_0 + \epsilon$ where $\epsilon$ is arbitrarily small is, to leading order (exact in the limit $\epsilon \to 0$)
\begin{align}
    G_i(t_0) = 
    \begin{cases}
    1 \text{ if } i=i_0\\
    0 \text{ if } i \neq i_0\\
    \end{cases}.
\end{align}

Thus, for $t>t_0$
\begin{align}
    \frac{dG_i}{dt} = (\mu + \alpha) \big( (i+1) G_{i+1} - iG_i \big),
\end{align}
with initial condition given by Equation (19). This is again the pure Markovian death process, so using Equation (5) we solve Equation (20) to give
\begin{align}
     G_i(t, i_0, t_0) = 
    \begin{cases}
    K(i, t, i_0, t_0, \mu + \alpha) \text{ if } t>t_0\\
    0 \text{ if } t<t_0\\
    \end{cases}.
\end{align}

Noting that $K(i, t, i_0, t_0, \mu + \alpha)$ is zero if $i>i_0$, the solution for $H_i$ becomes
\begin{align}
    H_i(t) = \sum^{N}_{i_0=i} \int^{t}_{0} \delta L^{k}_{i_0}(t_0) K(i, t, i_0, t_0, \mu+\alpha) dt_0.
\end{align}

Substituting Equation (15) into Equation (22), we find that
\begin{align}
    H_i(t) = \int^{t}_{0} \frac{\delta^k t_0^{k-1}}{(k-1)!} \exp{(-\delta t_0)} \exp{(-N \mu t)} \exp{(-i_0 (\mu + \alpha) (t-t_0))} \times S dt_0,
\end{align}
where
\begin{align}
    S = \sum^{N}_{i_0=i} \genfrac(){0pt}{0}{N}{i} \genfrac(){0pt}{0}{i_0}{i}  \big[ \exp{(\mu t)} - 1 \big]^{N-i} \big[ 1 - \exp{(-(\mu + \alpha) (t-t_0))} ]^{i_0-i}. 
\end{align}

Noting the identity
\begin{align}
   \genfrac(){0pt}{0}{N}{i_0} \genfrac(){0pt}{0}{i_0}{i} = \genfrac(){0pt}{0}{N-i}{i_0-i} \genfrac(){0pt}{0}{N}{i},
\end{align}
and substituting Equation (25) into Equation (24), we find
\begin{align}
    S = \genfrac(){0pt}{0}{N}{i} \sum^{N}_{i_0=i} \genfrac(){0pt}{0}{N-i}{i_0-i} \big[ \exp{(\mu t_0)} - 1 \big]^{N-i} \big[ 1 - \exp{(-(\mu + \alpha) (t-t_0))} ]^{i_0-i}.
\end{align}

Changing the index of summation $i' = i_0 - i$, we obtain
\begin{align}
    S = \genfrac(){0pt}{0}{N}{i} \sum^{N-i}_{i'=0} \genfrac(){0pt}{0}{N-i}{i'} \big[ \exp{(\mu t_0)} - 1 \big]^{N-i-i'} \big[ 1 - \exp{(-(\mu + \alpha) (t-t_0))} ]^{i'}.
\end{align}

Applying the Binomial formula, we can now write
\begin{align}
    S = \genfrac(){0pt}{0}{N}{i} \big[ \exp{(\mu t_0)} - \exp{(-(\mu + \alpha) (t-t_0))} \big]^{N-i}.
\end{align}

Substituting Equation (28) into (23) yields
\begin{align}
    H_i(t) = \int^{t}_{0} \genfrac(){0pt}{0}{N}{i} \frac{\delta^k t_0^{k-1}}{(k-1)!} \exp{(-\delta t_0)} \frac{\big[ 1 - \exp{(-\mu t_0 - (\mu + \alpha)(t-t_0))} \big]^N}{\big[ \exp{(\mu t_0 + (\mu + \alpha)(t-t_0))} - 1 \big]^i} dt_0.
\end{align}

\vspace{2mm}

\subsection{Expected Number of Nonlatent Hypnozoites}

Let $H^{\text{White}}(t)$ denote the number of nonlatent hypnozoites at time $t$ after inoculation. For notational convenience, we define
\begin{align}
    \exp{(\mu t_0 - (\mu + \alpha)(t-t_0))} = e^{c(t, t_0)}.
\end{align}

We now compute the expected number of nonlatent hypnozoites time $t$ after inoculation, as predicted by White's model
\begin{align}
    \mathbb{E}[H^{\text{White}}(t)] & = \sum^{N}_{i=1} i H_i(t) = \int^{t}_{0}  \frac{\delta^k t_0^{k-1}}{(k-1)!} e^{-\delta t_0} \big[ 1 - e^{-c(t, t_0)} \big]^N\ \sum^{N}_{i=1} i \genfrac(){0pt}{0}{N}{i} \frac{1}{\big[ e^{c(t, t_0)} - 1 \big]^i} dt_0.
\end{align}

Noting the identity
\begin{align}
    i \genfrac(){0pt}{0}{N}{i} = N \genfrac(){0pt}{0}{N-1}{i-1},
\end{align}
and substituting Equation (32) into Equation (31), we find
\begin{align}
    \mathbb{E}[H^{\text{White}}(t)] & = \int^{t}_{0}  \frac{\delta^k t_0^{k-1}}{(k-1)!} e^{-\delta t_0} \frac{N \big[ 1 - e^{-c(t, t_0)} \big]^N}{\big[ e^{c(t, t_0)} - 1 \big]} \sum^{N}_{i=1} \genfrac(){0pt}{0}{N-1}{i-1}  \frac{1}{\big[ e^{c(t, t_0)} - 1 \big]^{i-1}} dt_0 \nonumber \\
    & = \int^{t}_{0}  \frac{\delta^k t_0^{k-1}}{(k-1)!} e^{-\delta t_0} \frac{N \big[ 1 - e^{-c(t, t_0)} \big]^N}{\big[ e^{c(t, t_0)} - 1 \big]} \Big( 1 + \frac{1}{\big[ e^{c(t, t_0)} - 1 \big]} \Big)^{N-1} dt_0 \nonumber \\
    & = \int^{t}_{0}  \frac{\delta^k t_0^{k-1}}{(k-1)!} \exp{(-\delta t_0)} N e^{-c(t, t_0)} dt_0,
\end{align}
where we have used the Binomial formula to simplify Equation (33).\\

Substituting Equation (30) into Equation (33) yields
\begin{align}
    \mathbb{E}[H^{\text{White}}(t)] & = N \frac{\delta^k}{(k-1)!} \exp{(-(\mu + \alpha)t)} \int^t_0 {t_0}^{k-1} \exp{((\alpha-\delta)t_0)} dt_0.
\end{align}

Substituting Equation (8) from the main text into Equation (34), we find that
\begin{align}
    \mathbb{E}[H^{\text{White}}(t)] & = N p_{\text{nonlatent}}(t),
\end{align}
where $p_{\text{nonlatent}}(t)$ is the probability of a single hypnozoite being non-latent time $t$ after inoculation, as determined from our model.

\printbibliography